\documentclass[sigconf,authorversion]{acmart}

\usepackage{xcolor}

\usepackage[ruled]{algorithm2e}
\usepackage{multirow}

\usepackage{amssymb,amsbsy}
\usepackage{bbm}
\usepackage{paralist}
\usepackage{placeins}
\usepackage{tikz}
\usepackage{pgfplots}
\usepackage{pgfplotstable}
\usepackage{balance}
\usepackage{hyphenat} 
\pgfplotsset{compat=newest,scaled y ticks=false}
\usepgfplotslibrary{groupplots}
\usetikzlibrary{shapes}
\pgfplotsset{scaled y ticks=false,
  yticklabel style={
        /pgf/number format/fixed,
        /pgf/number format/precision=4
  },
}

\def\BibTeX{{\rm B\kern-.05em{\sc i\kern-.025em b}\kern-.08emT\kern-.1667em\lower.7ex\hbox{E}\kern-.125emX}}

\newcommand{\KL}[2]{\mathrm{KL}\left(#1 \middle\| #2\right)}

\def\x{{\bf x}}
\def\XX{{\bf X}}

\def\bZero{{\bf 0}}

\def\RR{{\bf R}}

\def\bw{{\bf w}}

\def\bp{{\bf p}}

\def\bm{{\bf m}}
\def\bz{{\bf z}}

\def\by{{\bf y}}

\def\bI{{\bf I}}

\def\N{{\mathcal N}}

\def\Mult{\mathrm{Mult}}
\def\bb{{\bf b}}
\def\RR{{\mathbb R}}
\def\I{{\bf I}}
\def\dec{\mathrm{dec}}
\def\enc{\mathrm{enc}}

\def\btheta{{\boldsymbol{\theta}}}
\def\bmu{{\boldsymbol{\mu}}}
\def\bsigma{{\boldsymbol{\sigma}}}
\def\bphi{{\boldsymbol{\phi}}}
\def\bpi{{\boldsymbol{\pi}}}
\def\beps{{\boldsymbol{\epsilon}}}

\def\bz{\boldsymbol{z}}
\def\bOne{\boldsymbol{1}}
\def\bw{\boldsymbol{w}}
\def\bx{\boldsymbol{x}}

\def\tbx{{\tilde \bx}}

\copyrightyear{2020}
\acmYear{2020}
\setcopyright{acmlicensed}
\acmConference[WSDM '20]{The Thirteenth ACM International Conference on Web Search and Data Mining}{February 3--7, 2020}{Houston, TX, USA}
\acmBooktitle{The Thirteenth ACM International Conference on Web Search and Data Mining (WSDM '20), February 3--7, 2020, Houston, TX, USA}
\acmPrice{15.00}
\acmDOI{10.1145/3336191.3371831}
\acmISBN{978-1-4503-6822-3/20/02}

\begin{document}

\title{RecVAE: a New Variational Autoencoder for Top-N Recommendations with Implicit Feedback}

\author{Ilya Shenbin}
\email{ilya.shenbin@gmail.com}
\affiliation{
  \institution{Samsung-PDMI Joint AI Center, Steklov Institute of Mathematics at St.~Petersburg, Russia}
}

\author{Anton Alekseev}
\email{anton.m.alexeyev@gmail.com}
\affiliation{
  \institution{Samsung-PDMI Joint AI Center, Steklov Institute of Mathematics at St.~Petersburg, Russia}
}

\author{Elena Tutubalina}
\email{tutubalinaev@gmail.com}
\affiliation{
  \institution{Samsung-PDMI Joint AI Center, Steklov Institute of Mathematics at St.~Petersburg, Russia}
}

\author{Valentin Malykh}
\email{valentin.malykh@phystech.edu}
\affiliation{
  \institution{Neural Systems and Deep Learning Laboratory, Moscow Institute of Physics and Technology, Russia}
}

\author{Sergey I. Nikolenko}
\email{sergey@logic.pdmi.ras.ru}
\affiliation{
  \institution{Samsung-PDMI Joint AI Center, Steklov Institute of Mathematics at St.~Petersburg, Russia}
}

\renewcommand{\shortauthors}{Shenbin et al.}

\begin{abstract}
Recent research has shown the advantages of using autoencoders based on deep neural networks for collaborative filtering. In particular, the recently proposed \emph{Mult-VAE} model, which used the multinomial likelihood variational autoencoders, has shown excellent results for top-N recommendations. In this work, we propose the \emph{Recommender VAE} (RecVAE) model that originates from our research on regularization techniques for variational autoencoders. \emph{RecVAE} introduces several novel ideas to improve \emph{Mult-VAE}, including a novel composite prior distribution for the latent codes, a new approach to setting the $\beta$ hyperparameter for the $\beta$-VAE framework, and a new approach to training based on alternating updates. In experimental evaluation, we show that \emph{RecVAE} significantly outperforms previously proposed autoencoder-based models, including \emph{Mult-VAE} and \emph{RaCT}, across classical collaborative filtering datasets, and present a detailed ablation study to assess our new developments. Code and models are available at \urlstyle{tt}\url{https://github.com/ilya-shenbin/RecVAE}.
\end{abstract}



\maketitle

\section{Introduction}

Matrix factorization (\emph{MF}) has become the industry standard as the foundation of recommender systems based on collaborative filtering. However, there are certain general issues that arise with this family of models. First, the number of parameters in any matrix factorization model is huge: it linearly depends on the number of both users and items, which leads to slow model learning and overfitting. Second, to make a prediction for a new user/item based on their ratings, one has to run an optimization procedure in order to find the corresponding user/item embedding. Third, only a small amount of ratings are known for some (often for a majority of) users and items, which could also lead to overfitting. This makes it necessary to heavily regularize matrix factorization models, and standard $L_1$ or $L_2$ regularizers are hard to tune.

Recently proposed models such as the \emph{Collaborative Denoising Autoencoder} (\emph{CDAE})~\cite{wu2016collaborative} partially solve these issues by using a parameterized function which maps user feedback to user embeddings. It performs regularization in an alternative way and makes it possible to predict item ratings for new users without additional iterative training. The \emph{Variational Autoencoder for Collaborative Filtering} (Mult-VAE)~\cite{liang2018variational} is a subsequent improvement of \emph{CDAE} that extends it to multinomial distributions in the likelihood, which are more suitable for recommendations.

In this work, we propose the \emph{Recommender VAE} (RecVAE) model for collaborative filtering with implicit feedback based on the variational autoencoder (VAE) and specifically on the \emph{Mult-VAE} approach. \emph{RecVAE} presents a number of important novelties that together combine into significantly improved performance. First, we have designed a new architecture for the encoder network. Second, we have introduced a novel composite prior distribution for the latent code $\bz$ in the variational autoencoder. The composite prior is a mixture of a standard Gaussian prior and the latent code distribution with parameters fixed from the previous iteration of the model (Section~\ref{sec:prior}), an idea originating from reinforcement learning where it is used to stabilize training~\cite{DBLP:journals/corr/SchulmanWDRK17,houthooft2016vime}. In the context of recommendations, we have also found that this prior improves training stability and performance. Third, we have developed a new approach to setting the hyperparameter $\beta$ for the Kullback-Leibler term in the objective function (Section~\ref{sec:kl}). We have found that $\beta$ should be user-specific, $\beta=\beta(\x_u)$, and should depend on the amount of data (implicit feedback) available for a given user.

Finally, we introduce a novel approach for training the model. In \emph{RecVAE}, training is done by alternating updates for the encoder and decoder (see Section~\ref{sec:als}). This approach has two important advantages. First, it allows to perform multiple updates of the encoder (a more complex network) for every update of the decoder (a very simple, single-layer network that contains item embeddings and biases). Second, it allows to use corrupted inputs (following the general idea of denoising autoencoders~\cite{im2017denoising,shu2018amortized}) only for training the encoder while still training the decoder on clean input data. This is again beneficial for the final model training due to the differing complexities of the encoder and decoder. 

As a result of the above novelties, our model significantly outperforms all autoencoder-based previous works and shows competitive or better results in comparison with other models across a variety of collaborative filtering datasets, including \emph{MovieLens-20M} (ML-20M), \emph{Netflix Prize Dataset}, and \emph{Million Songs Dataset} (MSD).

The paper is organized as follows. In Section~\ref{sec:background} we review the crucial components and approaches we are to employ in the proposed methods as well as other relevant prior art.
In Section~\ref{sec:method} we describe the basic \emph{Mult-VAE} approach and our modifications. Section~\ref{sec:experiments} contains the results of a comprehensive experimental study for our model, 
and Section~\ref{sec:conclusion} concludes the paper.

\section{Background and related work}\label{sec:background}

\subsection{Variational autoencoders and their extensions}\label{sec:vae}

The \emph{variational autoencoder} (VAE)~\cite{DBLP:journals/corr/KingmaW13,DBLP:conf/icml/RezendeMW14} is a deep latent variable model able to learn complex distributions. We begin with a brief exposition of the basic assumptions behind VAE that have been extended for collaborative filtering in \emph{Mult-VAE} and will be further extended in this work with \emph{RecVAE}. First, under the assumption that the dataset belongs to a low-dimensional manifold embedded in a high-dimensional space, the marginal likelihood function can be expressed via a latent code $\bz$ as $p_\theta(\bx) = \int p_\theta(\bx|\bz)p(\bz)d\bz$. Since the marginal likelihood function is intractable, it is usually approximated with the evidence lower bound (ELBO):
\begin{multline}\label{eq:elbo_vae}
    \log p_{\btheta}(\bx) \ge \mathcal{L}_{VAE} = \\
    \mathbb{E}_{q_{\bphi}(\bz|\bx)}\left[\log p_{\btheta}(\bx|\bz) - \KL{q_{\bphi}(\bz|\bx)}{p(\bz)}\right],
\end{multline}

\noindent
where $\mathrm{KL}$ is the Kullback-Leibler divergence, $p(\bz)$ is the prior distribution, $q_{\bphi}(\bz|\bx)$ is a variational approximation of the posterior distribution defined as a parameterized function with $\bphi$, and $\btheta$ are the parameters of $p_{\btheta}(\bx|\bz)$. This technique, known as \emph{amortized inference}, provides additional regularization and allows to obtain variational parameters with a closed form function. 

Variational autoencoders can be used not only as generative models but also for representation learning. $\beta$-VAE~\cite{higgins2017beta} is a modification of VAE designed to learn so-called \emph{disentangled representations} by adding a regularization coefficient to the Kullback-Leibler term in the evidence lower bound. The objective function of $\beta$-VAE can still be considered as a valid ELBO with additional approximate posterior regularization that rescales the Kullback-Leibler divergence in formula~\eqref{eq:elbo_vae} above by a factor of~$\beta$~\cite{hoffman2017beta,DBLP:conf/icml/MathieuRST19}:
\begin{equation}\label{eq:elbo_bvae}
    \mathcal{L}_{\beta\textrm{-}VAE} =
    \mathbb{E}_{q_{\bphi}(\bz|\bx)}\left[\log p_{\btheta}(\bx|\bz) - \beta\KL{q_{\bphi}(\bz|\bx)}{p(\bz)}\right].
\end{equation}

\emph{Denoising variational autoencoders} (DVAE)~\cite{im2017denoising,shu2018amortized}, similar to denoising autoencoders, are trying to reconstruct the input from its corrupted version. The ELBO in this model is defined as
\begin{multline}\label{eq:elbo_dvae}
    \log p_{\btheta}(\bx) \ge \mathcal{L}_{\mathrm{DVAE}} = \\
    \mathbb{E}_{q_{\bphi}(\bz|\tilde{\bx})}
    \mathbb{E}_{p(\tilde{\bx}|\bx)}\left[
    \log p_{\btheta}(\bx|\bz) - \KL{q_{\bphi}(\bz|\tilde{\bx})}{p(\bz)}\right].
\end{multline}

\noindent
It differs from VAE by the additional expectation $\mathbb{E}_{p(\tilde{\bx}|\bx)}$, where $p(\tilde{\bx}|\bx)$ is a noise distribution, usually Bernoulli or Gaussian. Similar to how denoising autoencoders are more robust and generalize better than regular autoencoders, DVAE improves over the performance of the basic VAE and makes it possible to learn more robust approximate posterior distributions.

The \emph{Conditional Variational Autoencoder} (CVAE)~\cite{sohn2015learning} is another VAE extension which is able to learn complex conditional distributions. Its ELBO is as follows:

\begin{multline}\label{eq:elbo_cvae}
    \log p_{\btheta}(\bx|\by) \ge \mathcal{L}_{CVAE} = \\
    \mathbb{E}_{q_{\bphi}(\bz|\bx, \by)}\left[\log p_{\btheta}(\bx|\bz, \by) - \KL{q_{\bphi}(\bz|\bx, \by)}{p_{\btheta}(\bz, \by)}\right],
\end{multline}
which is the same as the ELBO for regular VAE shown in~\eqref{eq:elbo_vae} but with all distributions conditioned on $\by$. \emph{VAE with Arbitrary Conditioning} (VAEAC) \cite{DBLP:conf/iclr/IvanovFV19}, based on CVAE, solves the imputation problem for missing features, which is in many ways similar to collaborative filtering. It has an ELBO similar to~\eqref{eq:elbo_cvae} where the variable are the unobserved features $\bx_b$ and the condition is $\bx_{1-b}$ for some binary mask $b$:
\begin{multline}\label{eq:elbo_vaeac}
    \log p_{\btheta,b}(\bx_b|\bx_{1-b}, b) \ge \\ \mathcal{L}_{VAEAC} = 
    \mathbb{E}_{q_{\bphi}(\bz|\bx, b)}\left[\log p_{\btheta}(\bx_b|\bz, \bx_{1-b}, b) - \vphantom{\KL{q_{\bphi}(\bz|\bx, b)}{p_{\btheta}(\bz\mid \bx_{1-b}, b)}} \right. \\
    \left. - \KL{q_{\bphi}(\bz|\bx, b)}{p_{\btheta}(\bz\mid \bx_{1-b}, b)}\right];
\end{multline}
an important point here is that the mask $b$ can be different and even have different number of ones (unobserved features) for different inputs $\x$. However, this approach cannot be directly applied to the implicit feedback case, which we consider in this work.

\subsection{Autoencoders and Regularization for Collaborative Filtering}\label{sec:ae_cf}

Let $U$ and $I$ be the sets of users and items respectively in a collaborative filtering problem. Consider the implicit feedback matrix $\mathbf{X} \in \{0, 1\}^{|U| \times |I|}$, where $x_{ui} = 1$ if the $u$th user positively interacted with (liked, bought, watched etc.) the $i$th item (movie, good, article etc.) and $x_{ui}=0$ otherwise. We denote by $\bx_u$ the feedback vector of user $u$.

The basic idea of the \emph{Collaborative Denoising Autoencoder} (\emph{CDAE}) model~\cite{wu2016collaborative} is to reconstruct the user feedback vector $\bx_u$ from its corrupted version $\tilde{\bx}_u$. The corrupted vector $\tilde{\bx}_u$ is obtained by randomly removing (setting to zero) some values in the vector~$\bx_u$. The encoder part of the model maps $\tilde{\bx}_u$ to the hidden state. In CDAE, both encoder and decoder are single neural layers (so the model itself is basically a feedforward neural network with one hidden layer), with a user input node providing user-specific weights and the rest of the weights shared across all users; the latent representation $\bz_u$ and reconstructed feedback $\hat{\bx}_u$ are computed as
\begin{align}
\bz_u &= \sigma\left(W^\top\tbx_u + V_u + \bb\right),\\
\hat{\bx}_u &= \sigma\left(W'\bz_u + \bb'\right),
\end{align}
where $W$ and $W'$ are input-to-hidden and hidden-to-output weight matrices respectively, and $V_u$ is the weight vector for the user input node). Note that the matrix $W$ can be considered as a matrix of item embeddings, and the matrix $V$, as the matrix of user embeddings.

Previous work also indicates that regularization plays a central role in collaborative filtering. Models based on matrix factorization (MF) with user/item embeddings almost invariably have an extremely large number of parameters that grows with dataset size; even a huge dataset with user ratings and other kinds of feedback cannot have more than a few dozen ratings per user (real users will not rate more items than that), so classical MF-based collaborative filtering models require heavy regularization~\cite{Park:2012:LRC:2181339.2181690,bell2007scalable,Adomavicius:2005:TNG:1070611.1070751}. The standard solution is to use simple $L_2$ or $L_1$ regularizers for the embedding weights, although more flexible priors have also been used in literature~\cite{salakhutdinov2008bayesian,lawrence2009non}. The works~\cite{wu2016collaborative,liang2018variational} present an alternative way of regularization based on the amortization of user embeddings coupled with denoising~\cite{vincent2010stacked,im2017denoising,shu2018amortized}. Several more models are reviewed in Section \ref{sec:baselines}.

The model which is nearest to our current work in prior art is \emph{Multinomial VAE} (\emph{Mult-VAE})~\cite{liang2018variational}, an extension of variational autoencoders for collaborative filtering with implicit feedback. In the next section, we begin with a detailed description of this model and then proceed to presenting our novel contributions.

\section{Proposed approach}\label{sec:method}

\subsection{Mult-VAE}\label{sec:mult}

We begin with a description of the \emph{Mult-VAE} model proposed in~\cite{liang2018variational}. The basic idea of \emph{Mult-VAE} is similar to VAE but with the multinomial distribution as the likelihood function instead of Gaussian and Bernoulli distributions commonly used in VAE. The generative model samples a $k$-dimensional latent representation $\bz_u$ for a user $u$, transforms it with a function $f_{\btheta}:\RR^k\to\RR^{|I|}$ parameterized by~$\btheta$, and then the feedback history $\bx_u$ of user $u$, which consists of $n_u$ interactions (clicks, purchases etc.), is assumed to be drawn from the multinomial distribution:
\begin{align}\label{eq:gen_multvae}
\bz_u &\sim \N(\bZero,\bI),\quad
\bpi(z_u) =\mathrm{softmax}(f_{\btheta}(\bz_u)),\\
\x_u &\sim \Mult(n_u,\pi(\bz_u)).
\end{align}
Note that classical collaborative filtering models also follow this scheme with a linear $f_{\btheta}$; the additional flexibility of \emph{Mult-VAE} comes from parameterizing $f$ with a neural network with parameters $\btheta$.

To estimate $\btheta$ one has to approximate the intractable posterior $p(\bz_u\mid\x_u)$. This, similar to regular VAE, is done by constructing an evidence lower bound for the variational approximation where $q(\bz_u)$ is assumed to be a fully factorized diagonal Gaussian distribution: $q(\bz_u) = \N\left(\bmu_u,\mathrm{diag}(\bsigma_u^2)\right)$. The resulting ELBO is

\begin{multline}\label{eq:elbo_multivae}
    \mathcal{L}_{\mathrm{Mult\text{-}VAE}} = \\ 
    \mathbb{E}_{q_{\bphi}(\bz_u|\bx_u)}
    \left[
    \log p_{\btheta}(\bx_u|\bz_u) - \beta \KL{q_{\bphi}(\bz_u|\bx_u)}{p(\bz_u)}\right],
\end{multline}
which follows the general VAE structure with an additional hyperparameter $\beta$ that allows to achieve a better balance between latent code independence and reconstruction accuracy, following the $\beta$-VAE framework~\cite{Higgins2017betaVAELB}.

The likelihood $p_{\btheta}(\bx_u|\bz_u)$ in the ELBO of \emph{Mult-VAE} is multinomial distribution. The logarithm of multinomial likelihood for a single user $u$ in \emph{Mult-VAE} is now
\begin{equation}
\log \Mult(\bx_u|n_u,\bp_u) = \sum_{i} \bx_{ui}\log \bp_{ui} + C_u,
\end{equation}
where $C_u$ is the logarithm of the normalizing constant which is ignored during training. We treat it as a sum of cross-entropy losses.

\subsection{Model Architecture}\label{sec:arch}

\begin{figure}[!t]
  \centering
  \includegraphics[width=.60\linewidth]{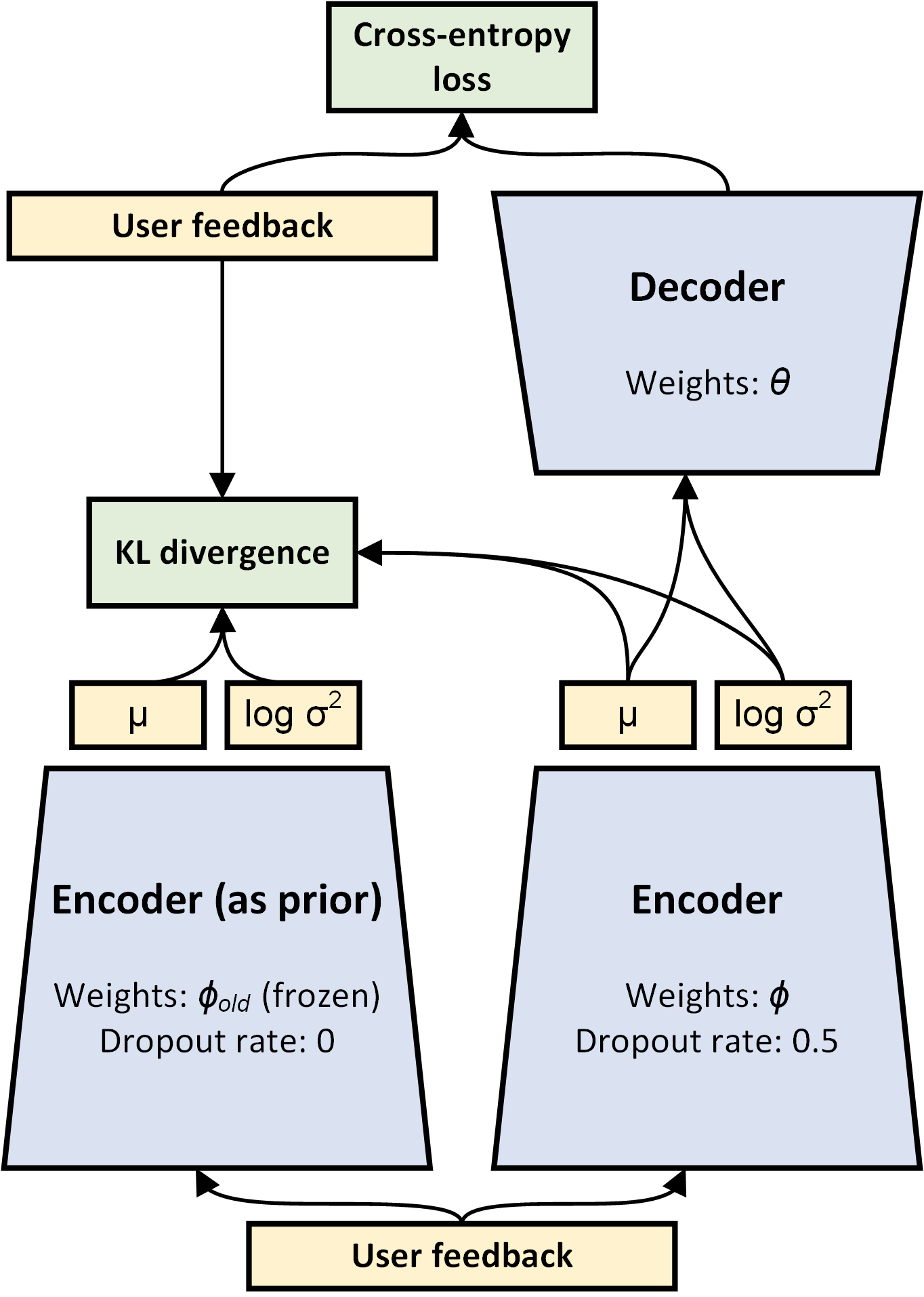}
  \caption{RecVAE architecture.}
  \label{fig:recvae}\vspace{-.45cm}
\end{figure}

Before introducing novel regularization techniques, we provide a general description of the proposed model. Our model is inherited from \emph{Mult-VAE}, but we also suggest some architecture changes. The general architecture is shown on Figure~\ref{fig:recvae}; the figure reflects some of the novelties we will discuss below in this section.

The first change is that we move to a denoising variational autoencoder, that is, move from the ELBO as shown in (\ref{eq:elbo_multivae}) to
\begin{multline}\label{eq:elbo_multdvae}
    \mathcal{L}_{\mathrm{Mult\text{-}VAE}} = 
    \mathbb{E}_{q_{\bphi}(\bz_u|\tbx_u)}
    \mathbb{E}_{p(\tbx_u|\bx_u)}
    \left[ \vphantom{- \beta \KL{q_{\bphi}(\bz_u|\bx_u)}{p(\bz_u)}}
    \log p_{\btheta}(\bx_u|\bz_u) - \right.
    \\ \left. - \beta \KL{q_{\bphi}(\bz_u|\bx_u)}{p(\bz_u)}\right].
\end{multline}
Note that while the original paper~\cite{liang2018variational} compares \emph{Mult-VAE} with \emph{Mult-DAE}, a denoising autoencoder that applies Bernoulli-based noise to the input but does not have the VAE structure (\emph{Mult-DAE} is a regular denoising autoencoder), in reality the authors used denoising for \emph{Mult-VAE} as well. This is evidenced both by their code base and by our experiments, where we were able to match the results of~\cite{liang2018variational} when we used denoising and got nowhere even close without denoising (we will return to this discussion in Section~\ref{sec:als}). According to our intuition and experiments, the input noise for the denoising autoencoder and latent variable noise of Monte Carlo integration play different roles: the former forces the model not only to reconstruct the input vector but also to predict unobserved feedback, while the latter leads to more robust embedding learning.


Similar to \emph{Mult-VAE}, we use the noise distribution $p(\tilde{\bx}|\bx)$ defined as elementwise multiplication of the vector $\bx$ by a vector of Bernoulli random variables parameterized by their mean $\mu_{\mathrm{noise}}$.

We keep the structure of both likelihood and approximate posterior unchanged: 
\begin{align}
p_{\btheta}(\bx_u|\bz_u) &= \mathrm{Mult}(\bx|n_u,\pi(\bz_u)), \\
\bpi(z_u) &=\mathrm{softmax}(f_{\btheta}(\bz_u)), \\
f_{\btheta}(\bz_u) &= W\bz_u+\bb, \\
q_{\bphi}(\bz_u|\bx_u) &= \mathcal{N}(\bz_u|\psi_\bphi(\bx_u)),
\end{align}
where $\psi_\bphi(\cdot)$ is the inference network, parameterized by $\bphi$, that predicts the mean vector and (diagonal) covariance matrix for the latent code $\bz_u$. However, we change the underlying neural networks. Our proposed architecture for the inference network is shown on Figure~\ref{fig:encoder}; it uses the ideas of densely connected layers from dense CNNs~\cite{DBLP:journals/corr/HuangLW16a}, swish activation functions~\cite{DBLP:conf/iclr/RamachandranZL18}, and layer normalization~\cite{2016arXiv160706450L}. The decoder network is a simple linear layer with softmax activation, where ${\btheta}=\{W, \bb\}$.
Here $W$ and $\bb$ can be considered as the item embeddings matrix and the item bias vector respectively. In a similar way, we can consider encoder $\psi_\bphi(\cdot)$ as a function that maps user feedback to user embeddings.

\subsection{Composite prior}\label{sec:prior}

Both input and output of \emph{Mult-VAE} are high-dimensional sparse vectors. Besides, while shared amortized approximate posterior regularizes learning, posterior updates for some parts of the observed data may hurt variational parameters corresponding to other parts of the data. These features may lead to instability during training, an effect similar to a well-known ``forgetting'' effect in reinforcement learning.  Previous work on policy-based reinforcement learning showed that it helps to regularize model parameters by bringing them closer to model parameters on the previous epoch~\cite{DBLP:journals/corr/SchulmanWDRK17,houthooft2016vime}; in reinforcement learning, it helps to make the final score grow more smoothly, preventing the model from forgetting good behaviours.

A direct counterpart of these ideas in our setting would be to use a standard Gaussian prior for the latent code $\bz$ and add a separate regularization term in the form of the KL divergence between the new parameter distribution $q_{\bphi}(\bz|\bx)$ and the previous one $q_{\bphi_{old}}(\bz|\bx)$, where $\bphi_{old}$ are the parameters from the previous epoch of the learning process. However, in our experiments it worked better to unite these two ideas (prior and additional regularizer) by using a composite prior
\begin{equation}\label{eq:composite_prior}
    p(\bz| \bphi_{old}, \bx ) = \alpha \mathcal{N}(\bz|0, \I) + (1-\alpha)q_{\bphi_{old}}(\bz|\bx),
\end{equation}
i.e., a convex combination (with $0 \leq \alpha \leq 1$) of a standard normal distribution and an approximate posterior $q_{\bphi_{old}}(\bz|\bx)$ with fixed parameters carried over from the previous epoch. The second term regulates large steps during variational parameters optimization and can be interpreted as an auxiliary loss function, while the first term prevents overfitting.

Note that this approach is not equivalent mathematically to a Gaussian prior and a separate KL regularizer that pulls current variational parameters to their previous values, and the fact that it works better makes this composite prior into a new meaningful contribution. We also note several works that argue for the benefits of trainable and/or complex prior distributions~\cite{tomczak2018vae,xu2019necessity}, although in our experiments these approaches have not brought any improvements compared to the prior proposed above.

Conditioning the prior on variational parameters from the previous training epoch converts our model to a conditional variational autoencoder where we assume both approximate posterior and likelihood to be conditionally independent of variational parameters from the previous epoch. Comparing our model to VAEAC~\cite{DBLP:conf/iclr/IvanovFV19}, the latter has a noised conditional prior while in our model we add noise to the approximate posterior during training. Also, unlike VAEAC, we do not train prior parameters.

\subsection{Rescaling KL divergence}\label{sec:kl}

We have already mentioned the $\beta$-VAE framework~\cite{Higgins2017betaVAELB} which is crucial for the performance of \emph{Mult-VAE} and, by extension, for \emph{RecVAE}. However, the question of how to choose or change $\beta$ is still not solved conclusively. Some works (see, e.g.,~\cite{bowman2016generating}) advocate to increase the value of $\beta$ from $0$ to $1$ during training, with $1$ yielding the basic VAE model and the actual ELBO. For the training of \emph{Mult-VAE}, the authors of~\cite{liang2018variational} proposed to increase $\beta$ from $0$ up to some constant. In our experiments, we have not found any improvements when $\beta$ is set to increase over some schedule, so we propose to keep scale factor fixed; this also makes it easier to find the optimal value for this hyperparameter.

\def\klu{\mathrm{KL}_u}
\def\klf{\mathrm{KL}_u^f}

Moreover, we propose an alternative view on KL divergence rescaling. Assume that the user feedback data is partially observed. We denote by $\XX_u^o$ the set of items which user $u$ has positively interacted with (according to observed data); $\XX_u^f$ similarly denotes the full set of items which user $u$ has positively interacted with (together with unobserved items). Items in $\XX_u^f$ and $\XX_u^o$ are represented in one-hot encoding, so that $\bx_u = \sum_{a \in \XX_u^o}\bOne_a$, where $\bx_u$ is the feedback vector for user $u$ and $\bOne_a$ is a vector with one $1$ in the position corresponding to item $a$. We denote 
$\bx_u = \sum_{a \in \XX_u^o}\bOne_a$, $\bx_u^f = \sum_{a \in \XX_u^f}\bOne_a$ and abbreviate 
$\klu = \KL{q_{\bphi}(\bz_u|\bx_u)}{p(\bz_u)}$ and $\klf = \KL{q_{\bphi}(\bz_u|\bx_u^f)}{p(\bz_u)}$.

Consider the evidence lower bound of a variational autoencoder~(\ref{eq:elbo_vae}) with multinomial likelihood. It can be rewritten as follows:

\begin{equation}\label{eq:1}
\begin{aligned}
\mathcal{L} &=
  \mathbb{E}_{q_{\bphi}(\bz_u|\bx_u^f)}\left[\log \mathrm{Mult}(\bx_u^f|\bpi(\bz_u)) - \klf\right] = \\
  & \mathbb{E}_{q_{\bphi}(\bz_u|\bx_u^f)} \left[\sum_{a \in \XX_u^f} \log \mathrm{Cat}(\bOne_a|\bpi(\bz_u)) - \klf\right] + C_u =\\
  & \mathbb{E}_{q_{\bphi}(\bz_u|\bx_u^f)} \sum_{a \in \XX_u^f} \left[\log \mathrm{Cat}(\bOne_a|\bpi(\bz_u)) - \frac{1}{|\XX_u^f|} \klf\right] + C_u,
\end{aligned}
\end{equation}
where $\mathrm{Cat}(\bOne_a|\bp_u) = p_{ua}$
is the categorical distribution,
and $C_u'$ a constant that depends on normalizing constant of the multinomial distribution $\mathrm{Mult}(\bx_u^f|\bpi(\bz_u))$, which does not affect optimization.

We approximate the ELBO obtained above by summing over observed feedback only and assuming that $q_{\bphi}(\bz_u|\bx_u) \approx q_{\bphi}(\bz_u|\bx_u^f)$ and therefore $\klu\approx\klf$. In the sequence of equalities below, we first approximate the sum over $\XX_u^f$ from~\eqref{eq:1} with a sum over $\XX_u^o$ with the corresponding rescaling coefficient ${|\XX_u^f|}/{|\XX_u^o|}$, then approximate $\klf$ with $\klu$:

\begin{equation*}
\begin{aligned}
\mathcal{L} \approx &
  \frac{|\XX_u^f|}{|\XX_u^o|} \mathbb{E}_{q_{\bphi}(\bz_u|\bx_u^f)} \sum_{a \in \XX_u^o} \left[\log \mathrm{Cat}(\bOne_a|\bpi(\bz_u)) - \frac{1}{|\XX_u^f|} \klf\right] + C'_u \\
  \approx & \frac{|\XX_u^f|}{|\XX_u^o|} \mathbb{E}_{q_{\bphi}(\bz_u|\bx_u)} \sum_{a \in \XX_u^o} \left[\log \mathrm{Cat}(\bOne_a|\bpi(\bz_u)) - \frac{1}{|\XX_u^f|} \klu\right] + C'_u \\
  = & \frac{|\XX_u^f|}{|\XX_u^o|}  \mathbb{E}_{q_{\bphi}(\bz_u|\bx_u)}\left[\sum_{a \in \XX_u^o} \log \mathrm{Cat}(\bOne_a|\bpi(\bz_u)) - \frac{|\XX_u^o|}{|\XX_u^f|} \klu\right] + C'_u \\
  = & \frac{|\XX_u^f|}{|\XX_u^o|}  \mathbb{E}_{q_{\bphi}(\bz_u|\bx_u)}\left[\log \mathrm{Mult}(\bx_u|\bpi(\bz_u)) - \frac{|\XX_u^o|}{|\XX_u^f|} \klu\right] + C''_u ,
\end{aligned}
\end{equation*}
where $C'_u$ and $C''_u$ are constants related to normalizing constants of the multinomial distribution.

Finally, we make the assumption that $|\XX_u^f|$ is the same for every $u$ and denote $\gamma={1}/{|\XX_u^f|}$. While this assumption might look strange, in effect $|\XX_u^f|$ is not merely unknown but is actually under our control: it is the number of items $u$ has feedback about plus the number of items for recommendation. Therefore, we reduce all $|\XX_u^f|$ into a single hyperparameter $\gamma$:
\begin{equation}\label{eq:3}
  \mathcal{L} \approx 
  \frac{1}{\gamma|\XX_u^o|}  \mathbb{E}_{q_{\bphi}(\bz_u|\bx_u)}\left[\log \mathrm{Mult}(\bx_u|\pi(\bz_u)) - \gamma|\XX_u^o| \klu\right].
\end{equation}

\noindent
In practice, we drop the ${1}/{\gamma|\XX_u^o|}$ factor in front of the expectation: it does not change the relation between the log likelihood and KL regularizer but rather changes the learning rate individually for each user, which slightly degraded performance in our experiments.
The resulting KL divergence scaling factor, 
\begin{equation}\label{eq:beta}
    \beta'=\beta'(\bx_u) = \gamma|\XX_u^o| = \gamma \sum_i x_{ui},
\end{equation}

\noindent
where $\gamma$ is a constant hyperparameter shared across all users and choosen with cross-validation, works better (see below). 
In total, we have proposed and motivated an approach where the $\beta$ constant in $\beta$-VAE is proportional to the amount of feedback available for the current user, $|\XX_u^o|$; this is an important modification that has led to significant improvements in our experiments.

\subsection{Alternating Training and Regularization by Denoising}\label{sec:als}

Alternating least squares (ALS)~\cite{bell2007scalable} is a popular technique for matrix factorization. We train our model in a similar way, alternating between user and item embeddings. User embeddings are amortized by the inference network, while each item embedding is trained individually. This means that the two groups of parameters, $\bphi$ in the encoder network and $\btheta$ in the item matrix and bias vector, are of a different nature and it might be best to train them in different ways. We propose to update $\bphi$ and $\btheta$ alternately with a different number of iterations: since the encoder is a much more complex network than the decoder, we make multiple updates of $\bphi$ for each update of $\btheta$.
This separation of training steps allows for another improvement. Both our experiments and prior art indicate that reconstruction of corrupted input data, i.e., using \emph{denoising} autoencoders, is necessary in autoencoder-based collaborative filtering, forcing the model to learn to not only reconstruct previously observed feedback but also generalize to unobserved feedback during inference. However, we noticed that performance improves if we do not corrupt input data during the training of $\btheta$ and leave the denoising purely for $\bphi$.
Since other types of regularization (such as $L_2$ or moving to a Bayesian decoder) also lead to degraded performance, it appears that decoder parameters are overregularized. Thus, we propose to train the decoder as part of a basic vanilla VAE, with no denoising applied. As for the encoder, however, we train it as part of the denoising variational autoencoder.

\subsection{Summary}

To summarize the proposed regularizers and changes, we first write down the ELBO for our model: 
\begin{multline}\label{eq:elbo}
    \mathcal{L} = 
    \mathbb{E}_{q_{\bphi}(\bz|\tilde{\bx})}
    \mathbb{E}_{q(\tilde{\bx}|\bx)}
    \left[
    \vphantom{\KL{q_{\bphi}(\bz|\tilde{\bx}}{p_{\bphi_{old}}(\bz|\bx)}}
    \log p_{\btheta}(\bx|\bz) - \right. \\
     \left. - \beta'(\bx)
    \KL{q_{\bphi}(\bz|\tilde{\bx})}{p(\bz| \bphi_{old}, \bx )}\right],
\end{multline}
where the conditional prior distribution $p(\bz| \bphi_{old}, \bx )$ has been defined in (\ref{eq:composite_prior}), and the modified weight $\beta'(\bx)$ of the KL term in (\ref{eq:beta}).

To keep the input uncorrupted while training the decoder, we introduce a modified objective function for $\btheta$ updates (we skip the KL term entirely because it does not depend on $\btheta$):

\begin{equation}\label{eq:elbo_dec}
    \mathcal{L}_{\dec} = \mathbb{E}_{q_{\bphi}(\bz|\bx)}     \log p_{\btheta}(\bx|\bz).
\end{equation}

\begin{algorithm}[t]
\caption{Proposed training procedure}\label{alg:trainer}
    \KwData{$\mathcal{D} = \{\bx_1, \dots, \bx_{|U|}\}$}
    \KwResult{$\bphi$, $\btheta$ }
    \For{$n := 1,...,N$}{
        \For{$m := 1,...,M_{\enc}$}{
        Sample batch $\{\bx_1, \dots, \bx_b\} \sim \mathcal{D}$\;
        Update $\bphi$ based on $\widetilde{\mathcal{L}}$\;
        }
        $\bphi_{old} := \bphi$\;
        \For{$m := 1,...,M_{\dec}$}{
        Sample batch $\{\bx_1, \dots, \bx_b\} \sim \mathcal{D}$\;
        Update $\btheta$ based on $\widetilde{\mathcal{L}}_{\dec}$\;
        }
    }
\end{algorithm}

We train the model using alternating updates as shown in Algorithm~\ref{alg:trainer}; different parameters $M_{\enc}$ and $M_{\dec}$ reflect the asymmetry between encoder and decoder that we have discussed above. During training, we approximate both inner and outer expectation by single Monte-Carlo samples and use reparametrization similar to regular VAE.
Now we can introduce the empirical lower bound 
\begin{multline}\label{eq:elbo_emp}
    \widetilde{\mathcal{L}}(\bx, \btheta, \bphi, \bphi_{old}) = 
    \log p_{\btheta}(\bx|\bz^{(*)}) -  \\
      - \beta'(\bx)
     \left(- \log q_{\bphi}\left(\bz^{(*)}|\tilde{\bx}\right) + \log p\left(\bz^{(*)}| \bphi_{old}, \bx\right)\right),
\end{multline}
where $\bz^{(*)} = g(\beps, \bmu, \bsigma)$, for a Gaussian posterior $g(\beps, \mu, \sigma) = \beps \cdot \bmu + \bsigma$, where $[\bmu, \log \bsigma^2]=\psi_{\bphi}(\x)$, $\beps \sim \N(\bZero,\bI)$, and $\tilde{\bx} = \bx \odot \bm$ is the noised input, $\bm \sim \mathrm{Bernoulli}(\mu_{\mathrm{noise}})$. We use Monte-Carlo sampling for both log likelihood and KL divergence since the KL divergence between a Gaussian and a mixture of Gaussians cannot be calculated analytically. The dropout layer on Figure~\ref{fig:encoder} serves for noising; it is turned off during evaluation, decoder learning phase, and in the composite prior. Since we use the multinomial likelihood, the main component of the loss function is the classification cross-entropy as shown on Figure~\ref{fig:recvae}. The empirical lower bound for decoder training $\widetilde{\mathcal{L}}_{\dec}$ is introduced in a similar way.

Similar to \emph{Mult-VAE}, our model is also able to make predictions for users whose feedback was not observed during training. To do that, we predict the user embedding with the inference network $\bz=\psi_{\bphi}(\x)$ based of the feedback $\x$ observed at test time and then predict top items using the trained decoder $p_{\btheta}(\x\mid\bz)$.

\section{Experimental evaluation}\label{sec:experiments}

\subsection{Metrics}

Following prior art, we evaluate the models with information retrieval metrics for ranking quality: Recall@$k$ and NDCG@$k$. Both metrics compares top-$k$ predictions of a model with the test set $\XX_u^t$ of user feedback for user $u$.
To obtain recommendations for \emph{RecVAE} and similar models, we sort the items in descending order of the likelihood predicted by the decoder, exclude items from the training set, and denote the item at the $n$th place in the resulting list as $R_u^{(n)}$. In this notation, evaluation metrics for a user $u$ are defined as follows:
\begin{equation}
\mathrm{Recall@}k(u) = \frac{1}{\min(M, |\XX_u^t|)}\sum_{n=1}^k {\mathbbm 1}\left[R_u^{(n)} \in \XX_u^t\right],
\end{equation}
where $\mathbbm{1}[\cdot]$ denotes the indicator function,

\begin{equation}
\begin{aligned}
\mathrm{DCG@}k(u) &= \sum_{n=1}^k \frac{2^{{\mathbbm 1}[R_u^{(n)} \in \XX_u^t]} - 1}{\log(n+1)},\\ 
\mathrm{NDCG@}k(u) &= \left(\sum_{n=1}^{|\XX_u^t|}\frac1{\log(n+1)}\right)^{-1} \mathrm{DCG@}k(u),
\end{aligned}
\end{equation}
i.e., $\mathrm{NDCG@}k(u)$ is defined as $\mathrm{DCG@}k(u)$ divided by its highest theoretically possible value.
Recall@$k$ accounts for all top-$k$ items equally, while NDCG@$k$ assigns larger weights to top ranked items; thus, it is natural to choose a larger value of $k$ for NDCG@$k$.

\subsection{Datasets}

We have evaluated \emph{RecVAE} on the \emph{MovieLens-20M} dataset\footnote{\url{https://grouplens.org/datasets/movielens/20m/}}~\cite{MovieLens}, \emph{Netflix Prize Dataset}\footnote{\url{https://www.netflixprize.com/}}~\cite{netflix}, and \emph{Million Songs Dataset}\footnote{\url{http://millionsongdataset.com/}}~\cite{songdata}.
We have preprocessed the datasets in accordance with the \emph{Mult-VAE} approach~\cite{liang2018variational}.
Dataset statistics after preprocessing are as follows:
\emph{MovieLens-20M} contains 9{,}990{,}682 ratings on 20{,}720 movies provided by 136{,}677 users,
\emph{Netflix Prize Dataset} contains 56{,}880{,}037 ratings on 17{,}769 movies provided by 463{,}435 users, and
\emph{Million Songs Dataset} contains 33{,}633{,}450  ratings on 41{,}140 songs provided by 571{,}355 users. 
In order to evaluate the model on users unavailable during the training, we have held out $10{,}000$ users for validation and testing for \emph{MovieLens-20M}, $40{,}000$ users for the \emph{Netflix Prize}, and $50{,}000$ users for the \emph{Million Songs Dataset}. We used $80\%$ of the ratings in the test set in order to compute user embeddings and evaluated the model on the remaining $20\%$ of the ratings.

\subsection{Baselines}\label{sec:baselines}

We compare the performance of the proposed model with several baselines, which we divide into three groups. 
The first group includes linear models from classical collaborative filtering. Weighted Matrix Factorization (WMF)~\cite{hu2008collaborative} binarizes implicit feedback $r_{ua}$ (number of times user $u$ positively interacted with item $a$) as $p_{ua}=\mathbbm{1}\left[r_{ua}>0\right]$ and decomposes the matrix of $p_{ua}$ similar to SVD but with confidence weights that increase with $r_{ua}$:
\begin{equation}
    \min_{\bw_u,\bw_a}\sum_{u,a}(1+\alpha r_{ua})\left(p_{ua}-\bw_u^\top\bw_a\right)^2 + \lambda(\|W_u\|_2 + \|W_a\|_2),
\end{equation}
where $\|\cdot\|_2$ is the $L_2$-norm.

The Sparse LInear Method (SLIM) for top-N recommendation~\cite{ning2011slim} learns a sparse matrix of aggregation coefficients $W$ that corresponds to the weights of rated items aggregated to produce recommendation scores, i.e., the prediction is ${\tilde r}_{ia}=\bw_i^\top\bw_a$, or in matrix form ${\tilde R} = RW$, with the resulting optimization problem
\begin{equation}
\min_W\frac12\left\|R-RW\right\|^2_F+\frac{\beta}2\|W\|^2_F+\lambda\|W\|_1
\end{equation}
subject to $W\ge 0$ and $\mathrm{diag}(W)=0$, where $\|\cdot\|_F$ is the Frobenius norm and $\|\cdot\|_1$ is the $L_1$-norm. The Embarrassingly Shallow Autoencoder (EASE)~\cite{steck2019embarrassingly} is a further improvement on SLIM with a closed-form solution, where the non-negativity constraint and $L_1$ regularization are dropped. Despite the name, we do not refer to this model as an autoencoder-based one.

The second is learning to rank methods such as WARP~\cite{weston2011wsabie} and \emph{LambdaNet}~\cite{burges2007learning}. 
\emph{LambdaNet} is a learning to rank approach that allows to work with objective functions that are either flat or discontinuous; the main idea is to approximate the gradient of each item's score in the ranked list for the corresponding query. The result can be treated as a direction where the item is to ``move'' in the ranked list for the query when sorted with respect to the newly predicted scores. WARP considers every user-item pair $(u, i)$ corresponding to a positive interaction when training to predict scores for recommendation ranking. For every user $u$, other random items $i'$ are sampled until the first one with the predicted score lower than that of the $i$ is found. The pair $(u, i')$ is then treated as a negative sample, and $(u, i)$ and $(u, i')$ are employed as positive and a negative contributions respectively for the approximation of the indicator function for the ranking loss similar to those introduced in an earlier work~\cite{usunier2009ranking}. For more details about WARP and its performance we refer to the original work~\cite{weston2011wsabie}.

The third group includes autoencoder-based methods that we have already discussed in Sections~\ref{sec:background} and~\ref{sec:mult}: CDAE~\cite{wu2016collaborative}, \emph{Mult-DAE}, and \emph{Mult-VAE}~\cite{liang2018variational}. 
The proposed \emph{RecVAE} model can also be considered as a member of this group.
We also consider the very recently proposed \emph{Ranking-Critical Training} (RaCT)~\cite{lobel2019towards}. This model adopts an actor-critic framework for collaborative filtering on implicit data. A critic (represented by a neural network) learns to approximate the ranking scores, which in turn improves an MLE-based nonlinear latent variable model (VAE and possibly its variations) with the learned ranking-critical objectives. The critic neural network is feature-based and is using posterior sampling as exploration for better estimates. Both actor and critic are pretrained in this model.
Scores for these models have been taken from~\cite{liang2018variational}~and~\cite{lobel2019towards}.

\subsection{Evaluation setup}

\emph{RecVAE} was trained with the \emph{Adam} optimizer~\cite{DBLP:journals/corr/KingmaB14} with learning~rate~$=5\cdot10^{-4}$, and batch size~$b=500$.
$M_{\mathrm{dec}}$ is selected so that each element in the dataset is selected once per epoch, i.e., $M_{\mathrm{dec}}=\frac{|U|}{\textrm{batch size}}$, the Bernoulli noise parameter is $\mu_{\mathrm{noise}}=0.5$, and $M_{\mathrm{enc}}=3M_{\mathrm{dec}}$. In addition to the standard normal distribution and the old posterior as parts of the composite prior, we also add a normal distribution with zero mean and $\log\sigma^2=10$ to the mixture. Weights of these mixture components are $3/20$, $3/4$, and $1/10$ respectively.
Since the model is sensitive to changes of the parameter $\gamma$, we have picked it individually for each dataset: $\gamma = 0.005$ for \emph{MovieLens-20M}, $\gamma = 0.0035$ for the \emph{Netflix Prize Dataset}, and $\gamma = 0.01$ for the \emph{Million Songs Dataset}. Each model was trained during $N=50$ epochs ($N=100$ for \emph{MSD}), choosing the best model by the NDCG@100 score on a validation subset. Initially, we fit the model for \emph{MovieLens-20M} and then fine-tuned it for the \emph{Netflix Prize Dataset} and \emph{MSD}.

\subsection{Results}

Performance scores for top-N recommendations in the three datasets are presented in Table~\ref{tab:vae-scores-last}. The results clearly show that in terms of recommendation quality, \emph{RecVAE} outperforms all previous autoencoder-based models across all datasets in the comparison, in particular, with a big improvement over \emph{Mult-VAE}. Moreover, new features of RecVAE and RaCT are independent and can be used together for even higher performance. Nevertheless, the proposed model significantly outperforms \emph{EASE} only on \emph{MovieLens-20M} datasets, and shows competitive performance on the \emph{Netflix Prize Dataset}.

For competing models in the comparison, we have used metrics reported in prior art; for \emph{RecVAE} we have also indicated confidence intervals, showing that the difference in scores is significant.

\begin{table}[!t]
\caption{Evaluation scores for \emph{RecVAE} and baseline models on \emph{MovieLens-20M}, \emph{Netflix Prize Dataset}, and MSD. The best results are highlighted in bold. The second best ones are underlined.} 
\centering
\begin{tabular}{l|l|l|l}\toprule
          & \bf Recall@20 & \bf Recall@50 & \bf NDCG@100 \\ \midrule
\multicolumn{4}{c}{\bf MovieLens-20M Dataset} \\\midrule
WARP     \cite{weston2011wsabie}       & 0.314 & 0.466 & 0.341 \\
LambdaNet\cite{burges2007learning}     & 0.395 & 0.534 & 0.427 \\
WMF      \cite{hu2008collaborative}    & 0.360 & 0.498 & 0.386 \\
SLIM     \cite{ning2011slim}           & 0.370 & 0.495 & 0.401 \\
CDAE     \cite{wu2016collaborative}    & 0.391 & 0.523 & 0.418 \\
Mult-DAE \cite{liang2018variational}   & 0.387 & 0.524 & 0.419 \\
Mult-VAE \cite{liang2018variational}   & 0.395 & 0.537 & 0.426 \\
RaCT     \cite{lobel2019towards}       & \underline{0.403} & \underline{0.543} & \underline{0.434} \\
EASE     \cite{steck2019embarrassingly}& 0.391 & 0.521 & 0.420 \\
RecVAE (ours)                          & \textbf{0.414}{\small$\pm$0.0027} & \textbf{0.553}{\small$\pm$0.0028} & \textbf{0.442}{\small$\pm$0.0021} \\ \midrule
\multicolumn{4}{c}{\bf Netflix Prize Dataset} \\\midrule
WARP     \cite{weston2011wsabie}       & 0.270 & 0.365 & 0.306 \\
LambdaNet\cite{burges2007learning}     & 0.352 & 0.441 & 0.386 \\
WMF      \cite{hu2008collaborative}    & 0.316 & 0.404 & 0.351 \\
SLIM     \cite{ning2011slim}           & 0.347 & 0.428 & 0.379 \\
CDAE     \cite{wu2016collaborative}    & 0.343 & 0.428 & 0.376 \\
Mult-DAE \cite{liang2018variational}   & 0.344 & 0.438 & 0.380 \\
Mult-VAE \cite{liang2018variational}   & 0.351 & 0.444 & 0.386 \\
RaCT     \cite{lobel2019towards}       & 0.357 & \underline{0.450} & 0.392 \\
EASE     \cite{steck2019embarrassingly}& \textbf{0.362} & 0.445 & \underline{0.393} \\
RecVAE (ours)                          & \underline{0.361}{\small$\pm$0.0013} & \textbf{0.452}{\small$\pm$0.0013} & \textbf{0.394}{\small$\pm$0.0010} \\ \midrule
\multicolumn{4}{c}{\bf Million Songs Dataset} \\\midrule
WARP     \cite{weston2011wsabie}       &  0.206 & 0.302 & 0.249 \\
LambdaNet\cite{burges2007learning}     &  0.259 & 0.355 & 0.308 \\
WMF      \cite{hu2008collaborative}    &  0.211 & 0.312 & 0.257 \\
SLIM     \cite{ning2011slim}           &  --    & --    & --    \\
CDAE     \cite{wu2016collaborative}    &  0.188 & 0.283 & 0.237 \\
Mult-DAE \cite{liang2018variational}   &  0.266 & 0.363 & 0.313 \\
Mult-VAE \cite{liang2018variational}   &  0.266 & 0.364 & 0.316 \\
RaCT     \cite{lobel2019towards}       &  0.268 & 0.364 & 0.319 \\
EASE     \cite{steck2019embarrassingly}&  \textbf{0.333} & \textbf{0.428} & \textbf{0.389} \\
RecVAE (ours)                          &  \underline{0.276}{\small$\pm$0.0010} & \underline{0.374}{\small$\pm$0.0011} & \underline{0.326}{\small$\pm$0.0010} \\ \bottomrule
\end{tabular}
\vspace{5pt}
\label{tab:vae-scores-last}\vspace{-.4cm}
\end{table}

\begin{figure}[!t]
    \centering
\captionof{table}{Evaluation of \emph{RecVAE} with different subsets of new features. The first row corresponds to \emph{Mult-VAE}.}
\begin{tabular}{lllll|p{1.3cm}p{1.3cm}p{1.3cm}}\toprule
\multirow{2}{*}[2.4cm]{\rotatebox{90}{\parbox{3.3cm}{\bf New architecture}}} &
\multirow{2}{*}[2.4cm]{\rotatebox{90}{\parbox{3.3cm}{\bf Composite prior}}} &
\multirow{2}{*}[2.4cm]{\rotatebox{90}{\parbox{3.3cm}{\bf $\beta(\bx)$ rescaling}}} &
\multirow{2}{*}[2.4cm]{\rotatebox{90}{\parbox{3.3cm}{\bf Alternating training}}} &
\multirow{2}{*}[2.4cm]{\rotatebox{90}{\parbox{3.3cm}{\bf Decoder w/o denoising}} }& 
\multicolumn{3}{c}{\bf NDCG@100 \rule{0pt}{2.7cm} }
 \\[.2cm] 
 & & & & & \bf ML-20M & \bf Netflix & \bf MSD  \\ \midrule
 &         &         &         &         & 0.426 & 0.386 & 0.319  \\ 
\checkmark&         &         &         &         & 0.428 & 0.388 & 0.320  \\ 
\checkmark&\checkmark&         &         &         & 0.435 & 0.392 & 0.325  \\ 
\checkmark&         &\checkmark&         &         & 0.435 & 0.390 & 0.321  \\ 
\checkmark&         &         &\checkmark&\checkmark& 0.427 & 0.387 & 0.319  \\ 
\checkmark&\checkmark&\checkmark&         &         & 0.438 & 0.390 & 0.325  \\ 
         &\checkmark&\checkmark&\checkmark&\checkmark& 0.420 & 0.380 & 0.308  \\ 
\checkmark&         &\checkmark&\checkmark&\checkmark& 0.434 & 0.383 & 0.321  \\ 
\checkmark&\checkmark&         &\checkmark&\checkmark& 0.437 & 0.392 & 0.323  \\ 
\checkmark&\checkmark&\checkmark&\checkmark&         & 0.441 & 0.391 & 0.322  \\ 
\checkmark&\checkmark&\checkmark&\checkmark&\checkmark& {\bf 0.442} & {\bf 0.394} & {\bf 0.326}  \\\bottomrule
\end{tabular}
\label{tab:analysis}\vspace{0.5cm}

    

    \input{fig_stability2}

    \captionof{figure}{Differences in NDCG@100 for a random user as a function of the training iteration.}
    \label{fig:embed}\vspace{-1.0cm}
\end{figure}

\subsection{Ablation study and negative results}

In order to demonstrate that each of the new features we introduced for \emph{RecVAE} compared to \emph{Mult-VAE} indeed helps to improve performance, we have performed a detailed ablation study, comparing various subsets of the features:
\begin{inparaenum}[(1)]
\item new encoder architecture,
\item composite prior for the latent codes,
\item $\beta$ rescaling,
\item alternating training, and 
\item removing denoising for the decoder.
\end{inparaenum}

Numerical results of the ablation study are presented in Table~\ref{tab:analysis}. We see that each new feature indeed improves the results, with all proposed new features leading to the best NDCG@100 scores on all three datasets. Some new features are complementary: e.g., $\beta$ rescaling and alternating training degrade the scores when applied individually, but together improve them; the new architecture does not bring much improvement by itself but facilitates other new features; $\beta$ rescaling is dataset-sensitive, sometimes improving a lot and sometimes doing virtually nothing.

We have also performed extended analysis of the composite prior, namely checked how the $\log p\left(\bz| \bphi_{old}, \bx\right)$ regularizer that brings variational parameters closer to old ones affects model stability. This regularizer stabilizes training, as evidenced by the rate of change in the variational parameters.
Figure~\ref{fig:embed} illustrates how the composite prior fixes the ``forgetting'' problem. It shows how NDCG@100 changes for a randomly chosen user as training progresses: each value is the difference in NDCG@100 for this user after each subsequent training update. Since each update changes the encoder network, it changes all user embeddings, and the changes can be detrimental for some users; note, however, that for the composite prior the changes remain positive almost everywhere while a simple Gaussian prior leads to much more volatile behaviour.

In addition, we would like to report the negative results of our other experiments. First, autoencoder-based models replace the matrix of user embeddings with a parameterized function, so it was natural to try to do the same for item embeddings. We trained \emph{RecVAE} with a symmetric autoencoder that predicts top users for a given item, training it alternately with regular \emph{RecVAE} and regularizing the results of each encoder with embeddings from the other model. The resulting model trained much slower, required more memory, and could not reach the results of \emph{RecVAE}.

Second, we have tried to use more complex prior distributions. Mixtures of Gaussians and the variational mixture VampPrior~\cite{tomczak2018vae} have (nearly) collapsed to a single node in our experiments, an effect previously noted in reinforcement learning~\cite{DBLP:journals/corr/abs-1809-10326}. The RealNVP prior~\cite{DBLP:conf/iclr/DinhSB17} has yielded better performance compared to the standard Gaussian prior, but we have not been able to successfully integrate it into the proposed composite prior: the composite prior with a Gaussian term remained the best throughout our experiments. We note this as a potential direction for further research.

Third, instead of $\beta$-VAE-like weighing of KL divergence, we tried to re-weigh each of the terms in the decomposed KL divergence separately. It appears natural to assume that ``more precise'' regularization could be beneficial for both performance and understanding. However, neither a simple decomposition into entropy and cross-entropy nor the more complex one proposed in \cite{chen2018isolating} has led the model to better results.

\section{Conclusion}\label{sec:conclusion}

In this work, we have presented a new model called \emph{RecVAE} that combines several improvements for the basic \emph{Mult-VAE} model, including a new encoder architecture, new composite prior distribution for the latent codes, new approach to setting the hyperparameter $\beta$, and a new approach to training \emph{RecVAE} with alternating updates of the encoder and decoder. As a result, performance of \emph{RecVAE} is comparable to EASE and significantly outperforms other models on classical collaborative filtering datasets such as \emph{MovieLens-20M}, \emph{Netflix Prize Dataset}, and \emph{Million Songs Dataset}.

We note that while we have provided certain theoretical motivations for our modifications, these motivations are sometimes incomplete, and some of our ideas have been primarily motivated by practical improvements. We believe that a comprehensive theoretical analysis of these ideas might prove fruitful for further advances, and we highlight this as an important direction for future research.

\FloatBarrier

\begin{acks}
This research was done at the Samsung-PDMI Joint AI Center at PDMI RAS and was supported by Samsung Research.
\end{acks}

\bibliographystyle{ACM-Reference-Format}
\bibliography{sample-base}

\appendix

\section{Network architecture}

\begin{figure}[h]
  \centering
  \includegraphics[width=.57\linewidth]{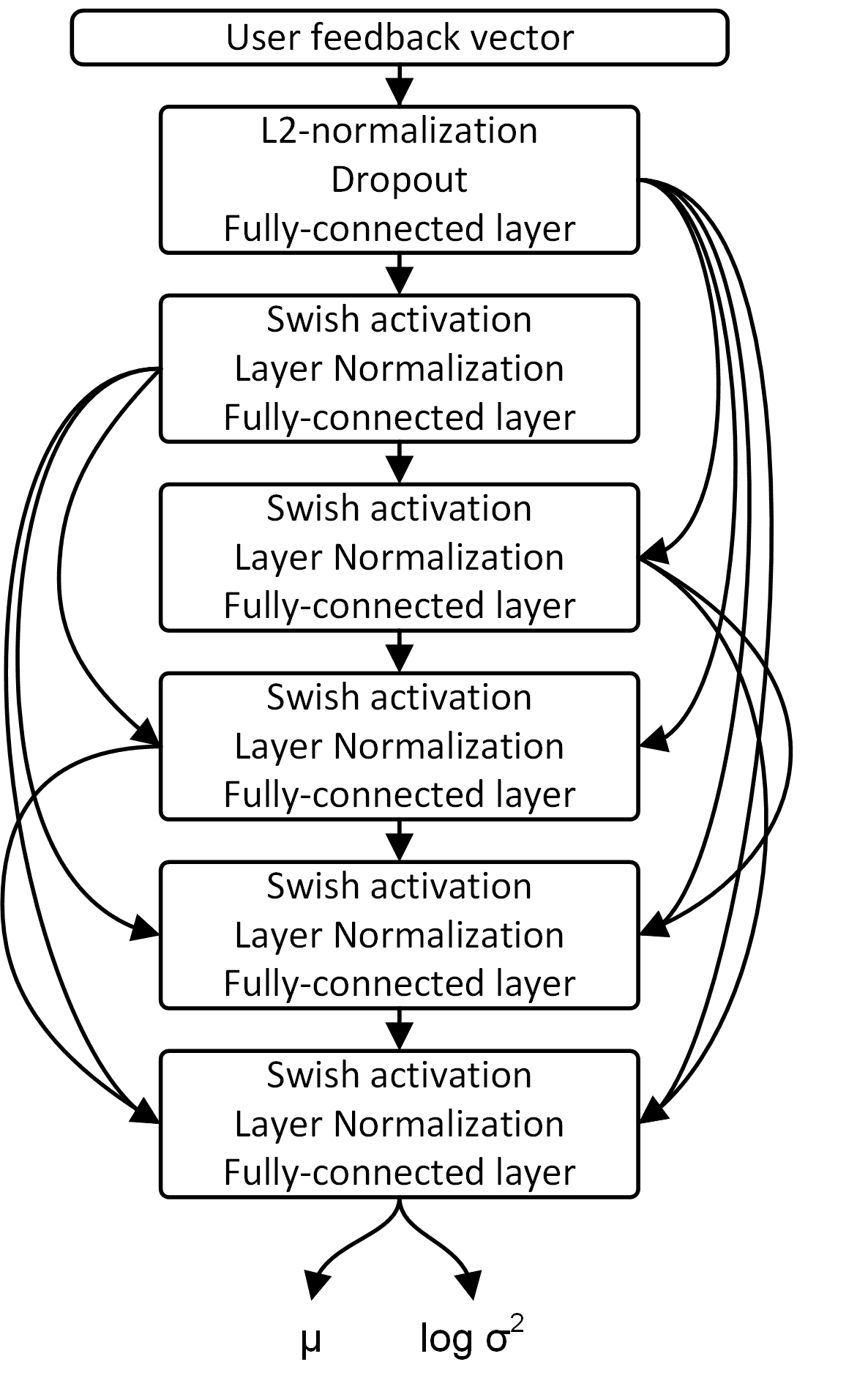}
  \caption{Architecture of the inference network $\psi_{\bphi}$.}
  \label{fig:encoder}
\end{figure}

\end{document}